\documentclass[journal=jpccck,manuscript=article]{achemso}
\setkeys{acs}{articletitle=true}
\usepackage{chemformula} 
\usepackage[T1]{fontenc} 
\usepackage{amssymb}
\usepackage{amsmath}

\usepackage{multirow}
\usepackage{booktabs}
\usepackage{subcaption}
\RequirePackage[normalem]{ulem}
\RequirePackage{color}\definecolor{RED}{rgb}{1,0,0}\definecolor{BLUE}{rgb}{0,0,1}

\author{Kangze Ren}
\affiliation[McMaster University]
{Department of Materials Science and Engineering, McMaster University, Hamilton, ON L8S 4L8, Canada}
\author{Chao Zheng}
\affiliation[McMaster University]
{Department of Materials Science and Engineering, McMaster University, Hamilton, ON L8S 4L8, Canada}
\author{Michael A. Brook}
\affiliation[McMaster University]
{Department of Chemistry  and Chemical Biology, McMaster University, Hamilton, ON L8S 4M1, Canada}
\author{Oleg Rubel}
\affiliation[McMaster University]
{Department of Materials Science and Engineering, McMaster University, Hamilton, ON L8S 4L8, Canada}
\email{rubelo@mcmaster.ca}

\title[Stability of Azr perovskite]
  {Stability of aziridinium lead iodide perovskite: ring strain and water vulnerability}


\begin{document}

\begin{tocentry}

\includegraphics{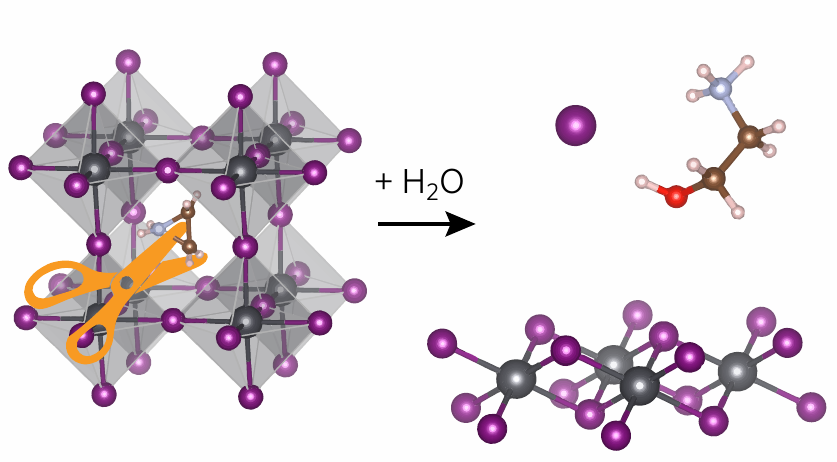}

\end{tocentry}

\begin{abstract}
Recently, an aziridinium lead iodide perovskite was proposed as a possible solar cell absorber material. We investigated the stability of this material using a density-functional theory with an emphasis on the ring strain associated with the three-membered aziridinium cation. It is shown that the aziridinium ring is prone to opening within the \ch{PbI3} environment. When exposed to moisture, aziridinium lead iodide can readily react with water. The resultant product will not likely be a stoichiometric lead halide perovskite structure.
\end{abstract}

\section{Introduction}

The power conversion efficiency of perovskite solar cell surpassed 24\% \cite{NREL-solar-cell-eff} only a decade after the first report of a perovskite solar cell \cite{Kojima_JACS_131_2009}. However, the lifetime of perovskite solar cells is still constrained by instability issues \cite{Meng_NC_9_2018}. Environmental factors such as moisture \cite{Niu_JMCA_2_2014}, ultraviolet light \cite{Sun_JMC_5_2017}, oxygen \cite{Bryant_EES_9_2016}, and thermal stress \cite{Mesquita_CSC_12_2019} are believed to be the major causes for degradation of perovskite solar cells. The vulnerability of perovskites is related to the intrinsic instability of lead halide perovskites \cite{Zhang_CPL_35_2018,Tenuta_SR_6_2016} and \ch{CH3NH3PbI3} in particular. This challenge inspired a search for intrinsically stable perovskite materials with promising photovoltaic properties.

The ionization energy of the cation on site $A$ in a general perovskite structure \ch{$AMX$3} can be used as one of indicators for predicting stability of hybrid halide perovskites \cite{Zheng_JPCC_121_2017}. Cations with a lower ionization energy tend to form more stable perovskite structures. \citeauthor{Zheng_JPCL_9_2018}\cite{Zheng_JPCL_9_2018} proposed a new perovskite material that features a three-membered ring cation, aziridinium \ch{CH2CH2N^{+}H2} (\ch{Azr+}) shown in Figs.~\ref{Fig:Structures}(a) and \ref{Fig:Diagrams}(a), with the ionization energy lower than methylammonium (MA). According to first-principle calculations \cite{Zheng_JPCL_9_2018,Teng_CPC_2018}, the decomposition reaction
\begin{equation}\label{Eq:Decomposition-Azr-1}
	\ch{AzrPbI3} \rightarrow \ch{AzrI} + \ch{PbI2}
\end{equation}
is energetically unfavorable. Thus, \ch{AzrPbI3} was proclaimed to be more stable than \ch{MAPbI3} \cite{Zheng_JPCL_9_2018}, which inspired follow up theoretical studies \cite{Teng_CPC_2018,Liu_JMCC_7_2019,Liu_JPCC_123_2019,Li_EES_12_2019}.

Although, on the theoretical level the perovskite structure of \ch{AzrPbI3} seems to be stable, the \ch{Azr+} ring itself is very volatile \cite{Bouyacoub_JMS_371_1996}, which raises a concern about feasibility of \ch{AzrPbI3} synthesis. In addition to volatility, according to \citeauthor{DeRycke_OL_13_2011} \cite{DeRycke_OL_13_2011}, a three-membered \ch{Azr+} ring is 17,000 times more reactive than a four-membered azetidinium ring (Fig.~\ref{Fig:Diagrams}). The high reactivity of \ch{Azr+} can be attributed to its high ring strain energy. Aziridine ring has a ring strain of 26.7~kcal/mol \cite{Dudev_JACS_120_1998}, which was measured with reference to the piperidine structure. A reason for the high strain energy in \ch{Azr+} is similar to that in cyclopropane. In the case of cyclopropane, the \ch{C-C-C} bond angle of $60^{\circ}$ is far from the ideal $109.5^{\circ}$ for bonds between atoms with $sp^3$ hybridized orbitals. A reduced hybridization between orbitals weakens the \ch{C-C} bonds making three-membered rings prone to opening \cite{Pearson_2010}. In the case of aziridine, the ring-opening reaction is even more favorable because amide anions are better leaving groups than carbanions \cite{Frank_2014}.

The neglect of the \ch{Azr+} ring-opening effect in our previous study \cite{Zheng_JPCL_9_2018} puts into question the feasibility of the proposed \ch{AzrPbI3} structure as a potential photovoltaic material. Here we use a density-functional theory (DFT) to further investigate stability of the \ch{Azr+} perovskite. We show that the \ch{Azr+} ring-opening and its reaction with water hinder synthesis of the previously proposed perovskite structure of \ch{AzrPbI3}.

\section{Method}

The closed-ring perovskite structure \ch{AzrPbI3} discussed in the current work is the same structure that was reported earlier \cite{Zheng_JPCL_9_2018}. The previously reported structure of the \ch{AzrI} salt \cite{Zheng_JPCL_9_2018} was created from the parent structure of formamidinium iodide \cite{Petrov_ACE_73_2017}. The organic compound \ch{CH3CHNH2} inside the open-ring \ch{AzrPbI3} was adopted from the most stable \ch{(C2H6N)+} isomorph\cite{Barone_JMS_124_1985}. The hexagonal structure of \ch{AzrPbI3} is inspired by the structure of $\delta$-\ch{HC(NH2)2PbI3} \cite{Stoumpos_IC_52_2013}. The crystal structure of 2-hydroxyethylammonium iodide was taken from Ref. \cite{Kohrt_ACE_70_2014}. The crystallographic information files (CIF files) with atomic structure of solids and molecules used in this work can be accessed through the Cambridge crystallographic data center (CCDC). \ch{AzrPbI3} closed-ring cubic, open-ring cubic, and open-ring hexagonal phases correspond to CCDC deposition numbers 1948430, 1947458, and 1947836, respectively. The structures of 2-iodoethanamine, a linear chain polyethylenimine, and 2-hydroxyethylammonium iodide can be found under CCDC deposition numbers 1948318, 1947839, and 1948421, respectively.

All of the electronic properties, as well as the total energy calculations have been accomplished in the framework of DFT \cite{Hohenberg_PR_136_1964,Kohn_PR_140_1965} with the Perdew-Burke-Ernzerhof \cite{Perdew_PRL_77_1996} (PBE) exchange-correlation functional. Dispersion interactions were incorporated at the DFT-D3 level\cite{Grimme_JCP_132_2010}. Calculations were performed with Vienna ab initio simulation program (VASP) and projector augmented-wave potentials \cite{Kresse_PRB_54_1996, Kresse_PRB_59_1999, Blochl_PRB_50_1994}. A conjugate-gradient algorithm was used to achieve ion relaxation by minimizing forces below 2~meV/{\AA} together with the cell shape and cell volume relaxation. Monopole, dipole and quadrupole corrections implemented in VASP were used to eliminate leading errors and acquiring accurate total energies of all charged ions\cite{Makov_PRB_51_1995, Neugebauer_PRB_46_1992}. The cut-off energy for a plane wave expansion was set at 500 eV. The total energy convergence was set at 0.1~$\mu$eV.

A $\Gamma$-centred grid was used to mesh the reciprocal space of solids. The mesh density was set at 25 points per 1~{\AA}$^{-1}$ length of their reciprocal lattice vectors. A single $\Gamma$ point was used for all gas-phase molecules.

\section{Results and Discussion}

\subsection{Aziridinium ring opening inside the \ch{PbI3} framework}

First we investigate the \ch{Azr+} ring stability within the \ch{PbI3} cage. It is assumed that the \ch{PbI3} cage deforms but remains intact as the \ch{Azr+} cation undergoes the ring-opening transformation. There are multiple possible open-ring \ch{Azr+} structures. \citeauthor{Barone_JMS_124_1985}\cite{Barone_JMS_124_1985} reported seven isomers as candidates for the \ch{Azr+} ring opening. Among those seven alternative structures, \ch{CH3CH=N^{+}H2} [shown in Figs.~\ref{Fig:Structures}(b) and \ref{Fig:Diagrams}(b)] possessed the lowest total energy based on molecular electronic structure calculations and was selected as the open-ring \ch{Azr+} structure in our study.

\begin{figure}[h]
  \includegraphics{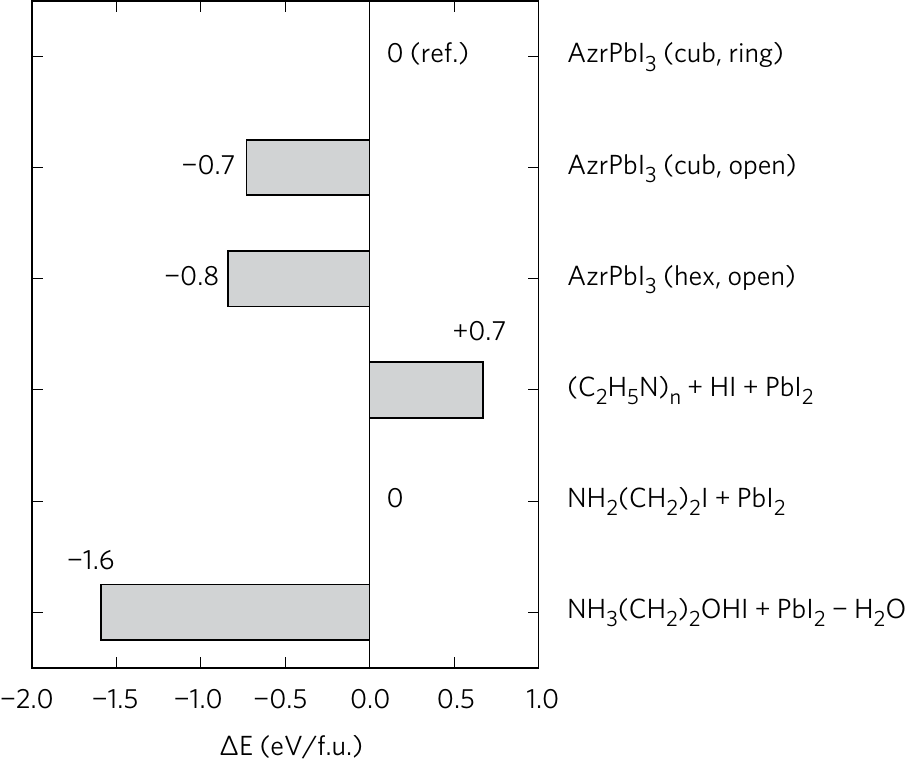}
  \caption{DFT total energies of various products associated with decomposition of \ch{AzrPbI3} taken relative to the closed-ring cubic phase. Negative values correspond to stable products. The chemical accuracy of DFT-PBE is about $\pm0.15$~eV/f.u.}
  \label{Fig:dE}
\end{figure}

The \ch{Azr+} ring strain within \ch{PbI3} framework was evaluated as the total energy difference
\begin{equation}\label{Eq:dE}
	\Delta E = E(\text{c-}\ch{AzrPbI3}) - E(\text{o-}\ch{AzrPbI3}),
\end{equation}
where $E(\text{c-}\ch{AzrPbI3})$ and $E(\text{o-}\ch{AzrPbI3})$ refer to the closed-ring and open-ring structures, respectively. The resultant $\Delta E$ is 0.7~eV per formula unit (f.u.) for the cubic phase (Fig.~\ref{Fig:dE}). (The cubic structure corresponds to a high temperature stable phase for hybrid halide perovskites\cite{Quarti_EES_9_2016,Whitfield_SR_6_2016}.) This calculated energy difference is less than the aziridine ring strain of 1.16~eV/f.u. reported in Ref.\cite{Dudev_JACS_120_1998} due to an additional chemical strain imposed on the cation by the \ch{PbI3} cage. The magnitude of $\Delta E$ is far beyond a PBE chemical uncertainty ($\pm 0.03$~eV/atom \cite{Hautier_PRB_85_2012}) and large enough to rule out existence of the closed-ring \ch{AzrPbI3} perovskite.

The next question is whether the open-ring perovskite structure of \ch{AzrPbI3} is stable and potentially useful for photovoltaics. The size of \ch{Azr+} in the open-ring configuration is relatively large. The hexagonal phase better accommodates organic cations of the size greater than MA. Both cubic and hexagonal phases of open-ring \ch{AzrPbI3} are shown in Fig.~\ref{Fig:Structures}(d,e). The hexagonal phase has a lower total energy by approximately 0.1~eV/f.u. (Fig.~\ref{Fig:dE}), which is within the DFT-PBE uncertainty and is of the same order as the final temperature corrections to the enthalpy and entropy of hybrid halide perovskites \cite{Tenuta_SR_6_2016}. The energy difference is, therefore, not conclusive to establish the room-temperature phase of  open-ring \ch{AzrPbI3} with certainty.

The band gap of open-ring cubic and hexagonal structures of \ch{AzrPbI3} was calculated at the DFT-PBE level without taking spin-orbit coupling into account. This approach yields a reliable estimate for the band gaps of halide perovskites due to cancelation of errors coming from the DFT semilocal exchange-correlation functional and omission of relativistic effects \cite{Even_JPCL_4_2013}. The open-ring cubic and hexagonal structure have the band gap of 1.6 and 2.1~eV, respectively. The band gap of the hexagonal structure is too far from the optimum range of $1.1-1.5$~eV for single-junction solar cells set by the Shockley-Queisser limit \cite{Shockley_JAP_32_1961}.

\subsection{Aziridinium ring opening without maintaining the \ch{PbI3} framework}

\begin{figure}[h]
  \includegraphics{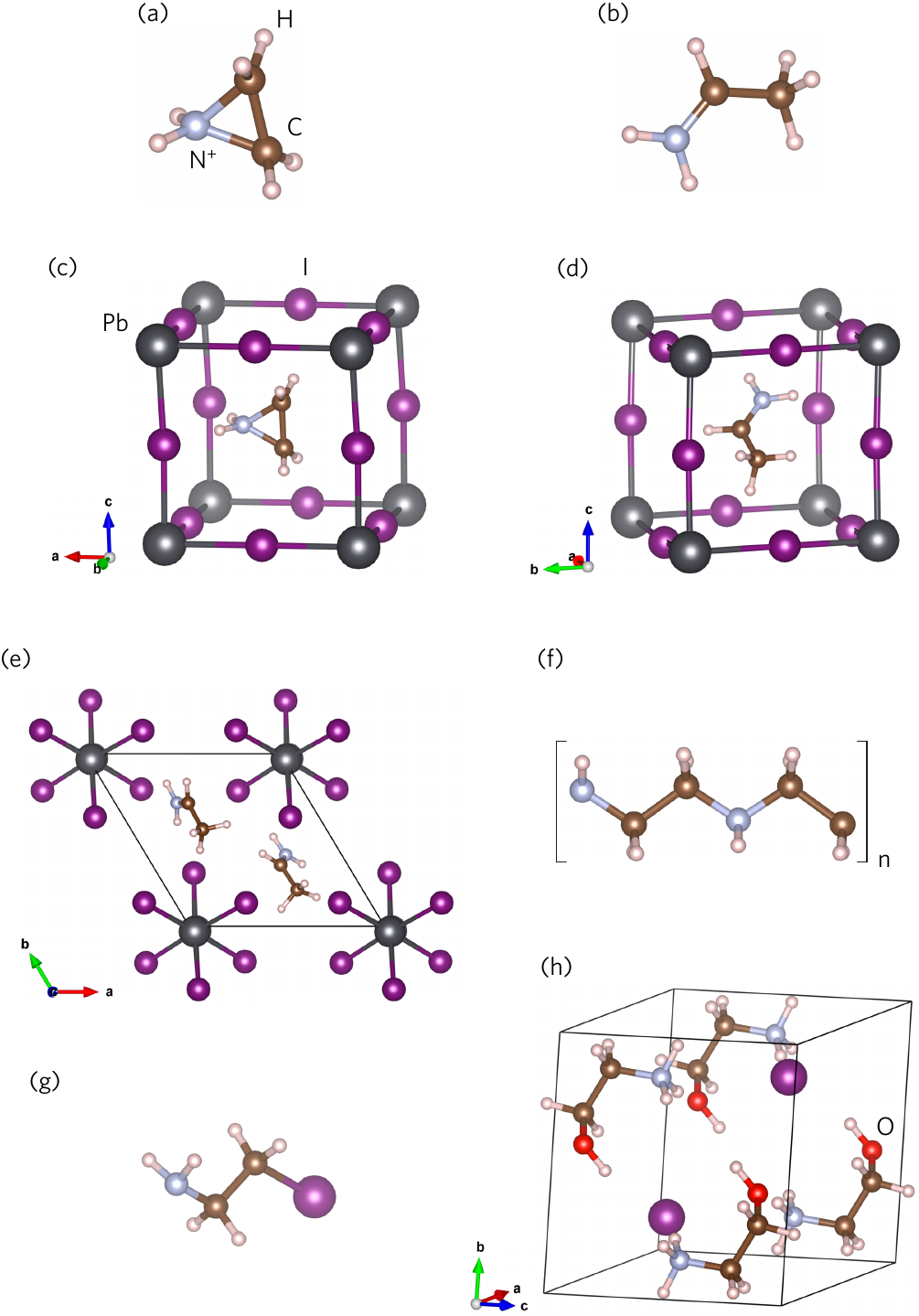}
  \caption{Structures used in total energy DFT calculations. (a) Three-membered ring \ch{Azr+}. (b) Most stable open-ring configuration for \ch{Azr+}. (c) Crystal structure of \ch{AzrPbI3} in a closed-ring cubic phase. (d) Open-ring cubic phase of  \ch{AzrPbI3}. (e) Open-ring hexagonal phase of  \ch{AzrPbI3}. (f) Linear chain polyethylenimine (\ch{C2H5N})$_n$. (g) 2-Iodoethanamine. (h) 2-Hydroxyethylammonium iodide.}
  \label{Fig:Structures}
\end{figure}

\begin{figure}[h]
  \includegraphics{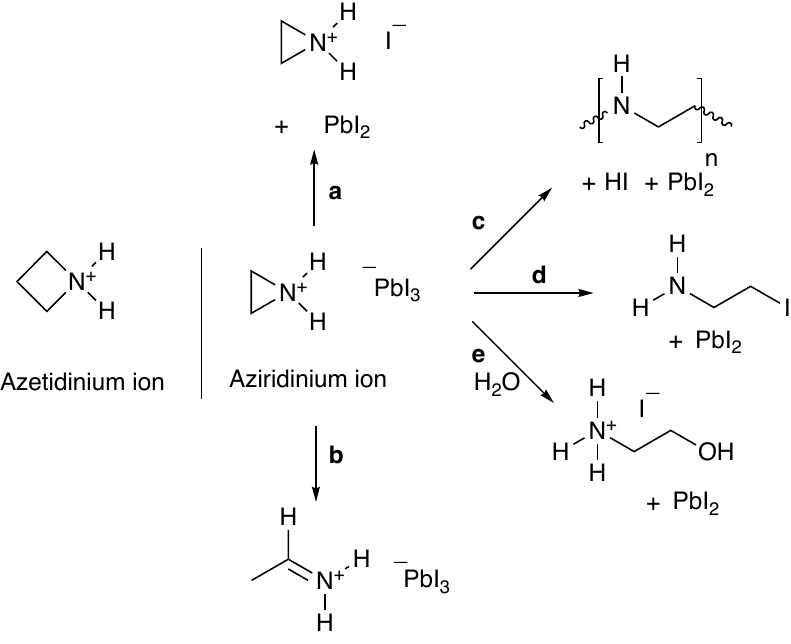}
  \caption{Structures of azetidinium ion and aziridinium ion; (a) Undergoing conversion of counterions. (b) Ring opening/rearrangement. (c) Polymerization to give polyethyleneimine. (d) Ring-opening nucleophilic attack by iodide to give iodoethylamine. (e) Hydrolysis to give the 2-hydroxyethylammonium iodide.}
  \label{Fig:Diagrams}
\end{figure}

Next we address stability of an open-ring \ch{AzrPbI3} phase. We investigate three decomposition pathways which feature the \ch{Azr+} ring opening and rupture of the \ch{PbI3} framework. The first structure is a linear polyethylenimine, the product of a suggested polymerization shown in Figs.~\ref{Fig:Structures}(f) and \ref{Fig:Diagrams}(c). The second structure is 2-iodoethanamine shown in Fig.~\ref{Fig:Structures}(g) and \ref{Fig:Diagrams}(d), generated under a halide nucleophilic ring-opening reaction. The third structure is 2-hydroxyethylammonium iodide shown in Fig.~\ref{Fig:Structures}(h) and \ref{Fig:Diagrams}(e), which is a water-induced product serving as the mean of testing the water resistance. The most favorable product, which features the lowest total energy, shall be determined via DFT calculations.

\subsubsection{Aziridinium polymerization}

Aziridine in acid medium generates \ch{Azr+} which undergoes polymerization with aziridine thanks to its minimum steric hindrance \cite{GOETHALS_ACS_1985} leading to a branched polyethylenimine as a product \cite{Gleede_PC_10_2019}. Due to the high similarity in chemical structure and similar thermodynamic/kinetic feasibility for polymerization with aziridine, we assume polymerized product of \ch{Azr+} is similar to the polymerized product of aziridine, a linear polyethylenimine.

Here, we choose to calculate the total of energy of linear polyethylenimine shown in Figs.~\ref{Fig:Structures}(f) and  \ref{Fig:Diagrams}(c) due to its simple structure and compatibility with periodic boundary conditions imposed by VASP. The linear polyethylenimine shares a similar chemical structure to branched polyethylenimine. The decomposition pathway is 
\begin{equation}\label{Eq:Decomposition-AzrPbI3-poly}
	\ch{AzrPbI3(s)} \rightarrow \ch{(C2H5N)}_n\text{(s)} + \ch{HI(g)} + \ch{PbI2(s)}
\end{equation}
where \ch{AzrPbI3(s)} refers to the hexagonal open-ring phase. The total energy of products in this reaction is 1.5~eV above the total energy of \ch{AzrPbI3(s)} (Fig.~\ref{Fig:dE}). This result rules out the polymer chain as a viable decomposition product.

\subsubsection{Halide nucleophilic attack}

The \ch{Azr+} ion can be considered as an activated aziridine, which easily undergoes ring-opening reaction depending on nature of nucleophiles \cite{Frank_2014}. This facile reaction is the origin of the biological activity of nitrogen mustards, which are active alkylating agents of biological molecules \cite{Singh_EJMC_151_2018}. \citeauthor{Cerichelli_JCSCC_1985}\cite{Cerichelli_JCSCC_1985} suggest that the protonation of nitrogen is potentially accountable for ring opening. Nucleophilic reactions with hydrogen protonated \ch{Azr+} were studied in Refs. \cite{Bouyacoub_JMS_371_1996, Dhooghe_EJOC_2010_2010} that feature a halide ion attacking on the C and liberating \ch{ICH2H2CNH2} from the ring. The resultant product 2-iodoethanamine (\ch{NH2(CH2)2I}) is inspired by structures presented in Refs. \cite{Bouyacoub_JMS_371_1996, Dhooghe_EJOC_2010_2010}. Several alternative arrangements of solid iodoethylamine were studied, and the lowest energy structure is shown in Figs.~\ref{Fig:Structures}(g) and \ref{Fig:Diagrams}(d).

The corresponding decomposition pathway can be expressed as
\begin{equation}\label{Eq:Decomposition-AzrPbI3-nuc}
	\ch{AzrPbI3(s)} \rightarrow \ch{NH2(CH2)2I(s)} + \ch{PbI2(s)}.
\end{equation}
The total energy difference of products in this reaction is 0.8~eV above the \ch{AzrPbI3} hexagonal open-ring phase (Fig.~\ref{Fig:dE}). This result shows 2-iodoethanamine is not a favorable product for the decomposition reaction. \citeauthor{Silva_TL_46_2005}\cite{Silva_TL_46_2005} have done a similar study on halide induced \ch{Azr+} ring opening. In their study, the energies of the three-membered ring and halide open-ring structures were very close. One group of data from Silva also shows that the closed-ring \ch{Azr+} halide is more stable than the open-ring \ch{Azr+} halide. In other words, \citeauthor{Silva_TL_46_2005}'s finding further confirms that the structure induced by nucleophilic ring-opening reaction is not as favorable as the ring rupturing in the \ch{PbI3} framework.

\subsubsection{Reaction with water}

It is critical to analyze degradation mechanisms that involve water for determining stability of halide perovskites. There are some ambiguities regarding the product of \ch{Azr+} reaction with water. \citeauthor{Bouyacoub_JMS_371_1996} \cite{Bouyacoub_JMS_371_1996} proposed \ch{H2N^{+}(CH2)2OH2}, whereas \citeauthor{Kohrt_ACE_70_2014} \cite{Kohrt_ACE_70_2014} suggested \ch{H3N^{+}(CH2)2OH}. The two structures differ by the location of a hydrogen atom. Our total energy calculations suggest the second structure as the most favorable product [Figs.~\ref{Fig:Structures}(h) and \ref{Fig:Diagrams}(e)]. 2-Hydroxyethylammonium (\ch{H3N^{+}(CH2)2OH}) is known to the perovskites community as a precursor for deficient perovskites \cite{Leblanc_ACIE_56_2017,Tsai_AEL_3_2018}. This cation is too large to form stoichiometric perovskite structures, but it can fill in for a \ch{(Pb-I)+} vacancy in perovskites.

The reaction with water can be expressed as
\begin{equation}\label{Eq:Decomposition-AzrPbI3-water}
	\ch{AzrPbI3(s)} + \ch{H2O(g)} \rightarrow \ch{NH3(CH2)2OHI(s)} + \ch{PbI2(s)}.
\end{equation}
The total energy difference for this reaction is 0.8~eV in favor of the right-hand side (Fig.~\ref{Fig:dE}). This result indicates that the open-ring \ch{AzrPbI3} is unstable in the moist environment.

\section{Conclusion}

Stability of a recently-proposed aziridinium lead iodide perovskite was investigated using density-functional theory. The work was motivated by a high ring strain associated with the three-membered aziridinium cation. We show that opening the aziridinium ring within the \ch{PbI3} cage liberates the energy of about $0.7-0.8$~eV per formula unit. This result rules out the possibility of an experimental synthesis of the aziridinium lead iodide perovskite material. In an open-ring configuration, the aziridinium lead iodide would likely adapt a hexagonal structure rather than a perovskite structure. However, the total energy difference between the two alternative structures is within the chemical uncertainty of an exchange-correlation functional used. The open-ring aziridinium cation is unstable in moist environments. It can readily react with water forming a 2-hydroxyethylammonium ion. Even though this cation is stable, it is too large for stoichiometric lead halide perovskite structures.





\begin{acknowledgement}
Authors are thankful to Prof.~Yurij Mozharivskyj (Department of Chemistry, McMaster University) for drawing our attention to instability of the three-membered ring structure. Authors also appreciate discussions with Dr.~James McNulty (Department of Chemistry and Chemical Biology, McMaster University) regarding the 2-iodoethanamine as one of the decomposed products. K.R. would like to acknowledge the financial support from the Natural Sciences and Engineering Research Council (NSERC) of Canada under the Undergraduate Student Research Awards program. M.A.B, C.Z., and O.R.  acknowledge funding provided by NSERC, the latter two under the Discovery Grant Program RGPIN-2015-04518. Calculations were performed using a Compute Canada infrastructure supported by the Canada Foundation for Innovation under the John R. Evans Leaders Fund program.
\end{acknowledgement}

\bibliography{../../bibliography}

\clearpage

\end{document}